# Nonspreading Light Pulses in Photonic Crystals

K.Staliunas[1], C.Serrat[2], R.Herrero[3], C.Cojocaru[2], and J.Trull[2]

[1]Institució Catalana de Reserca i Estudis Avançats (ICREA), Departament de Fisica i Enginyeria Nuclear, Universitat Politècnica de Catalunya, Colom 11, E-08222 Terrassa, Barcelona, Spain;
[2]Departament de Fisica i Enginyeria Nuclear, Universitat Politecnica de Catalunya, Colom 11, E-08222 Terrassa, Barcelona, Spain.
[3]Departament de Fisica i Enginyeria Nuclear, Universitat Politecnica de Catalunya, Comte Urgell 187, E-08036 Barcelona, Spain

We investigate propagation of light pulses in photonic crystals in the vicinity of the zero-diffraction point. We show that Gaussian pulses due to nonzero width of their spectrum spread weakly in space and time during the propagation. We also find the family of nonspreading pulses, propagating invariantly in the vicinity of the zero diffraction point of photonic crystals.

Since the initial proposal of the concept of Photonic Crystals (PC) in 1987 [1] a number of studies showed that these materials with a periodic modulation of the refraction index on a spatial scale of the order of the wavelength of light, are powerful tools to control and modify the propagation of electromagnetic fields. PCs, offer a possibility of engineering the dispersion properties of light to yield photonic band gaps in the transmission and reflection spectra, thus they can be designed to act as light conductors or insulators [2]. PCs modify (in particular reduce strongly) the phase and group velocities of light. This effect produces an increase of the photon lifetime within small distances and creates a strong localization of the electromagnetic field [3]. Recently it becomes apparent that the PCs can also modify the diffraction of light, in that the diffraction can become negative if the refractive index is modulated in direction perpendicular to the propagation of the light (1-dimensional PCs) [4]. Negative diffraction was also predicted for acoustic [5], and matter waves [6] in periodic materials.

If the diffraction is positive at one edge of propagation band, as is negative at the other edge, an inflection point inside of phonic band can be expected, characterized by the vanishing diffraction. The vanishing of the diffraction has been shown for the arrays of waveguides in [7], and more recently for 2- and 3-dimensional PCs [8]. The vanishing of diffraction means, that at some particular point of parameter space, and for a given frequency, the curvature of the spatial dispersion curve $1/2 \cdot \partial^2 k_{II}/\partial k_\perp^2$ becomes zero (here $k_{II}$ is the longitudinal- and $k_\perp$ is the transverse component of the wave-vector).

All the above studies consider diffraction management of monochromatic light beams. For pulses of nonzero width of the spectra, the diffraction can disappear for a particular frequency only. The other frequency components, not corresponding exactly to the nondiffractive point, broaden diffractively in propagation. This can result in a complicated shaping of the pulse propagating in the vicinity of the zero-diffraction point. In addition the presence of periodic modulation of refraction index introduces the group velocity dispersion [3], causing the temporal broadening of the pulses. Having in mind these two ingredients caused by PCs (diffraction dependence on the frequency, and appearance of group velocity dispersion), the pulse propagation in PCs becomes very complicated.

This letter is devoted to the study of the pulse propagation in the vicinity of the zero diffraction point. First we analyze the pulse propagation close to the zero diffraction point for PCs, and develop the normal form description close to this point. Next, basing on the normal form description, we analyze the pulse propagation: calculate numerically the evolution of the shape, and evaluate the (weak) spatio-temporal broadening close to the zero diffraction point. Finally we propose the families of the pulses with particular spatio-temporal form, which do not broaden in space and time during the propagation. Such nondiffractive pulses in PCs are similar to the X-pulses in dispersive and diffractive materials [9]. The X-pulses have particular envelopes of spatio-temporal spectra, such that the dephasing of different spatial components due to diffraction and dispersion mutually compensates, and the pulses propagate without broadening. Analogously the nonspreading pulses, as suggested by us, have specific envelopes of the spatio-temporal spectra in the vicinity of the zero diffraction point which mutually compensate dispersion and (weak) diffraction in propagation through PCs.

We consider two-dimensional PC, consisting of superposition of two periodic and harmonic (i.e. sinusoidal) lamellae-like refraction index gratings: $\Delta n(\mathbf{r}) = 2m(\cos(\mathbf{q}_1\mathbf{r}) + \cos(\mathbf{q}_2\mathbf{r}))$ with $|\mathbf{q}_1| = |\mathbf{q}_2| = q$ at angles $\pm\alpha$ to the optical axis. This results in: $\Delta n(x,z) = 4m\cos(q_\perp x)\cos(q_{II} z)$, with $q_{II} = q\cos(\alpha)$, and $q_\perp = q\sin(\alpha)$. The crystallographic axes of such PC are $\pi \cdot (\pm 1/q_\perp, 1/q_{II})$, and the reciprocal lattice vectors are $\mathbf{q}_1$ and $\mathbf{q}_2$. We describe the light propagation under approximation of slowly (in space and time) varying envelopes [10]:

$$\left(\frac{1}{c}\frac{\partial}{\partial t} + \frac{\partial}{\partial z} - \frac{i}{2k_0}\frac{\partial^2}{\partial x^2} - i\Delta n(x,z)k_0\right) A(x,z,t) = 0 \quad (1)$$

Here $A(x,z,t)$ is the complex envelope of the electromagnetic field in two-dimensional space $(x,z)$, and retarded time $t$: $E(x,z,t) = A(x,z,t)e^{ik_0 z - i\omega_0 t}$, propagating along the z-direction with a wave-number $k_0 = \omega_0/c$ We consider next the reference frame moving with the velocity of light, thus the first term of Eq.(1) disappears.

First we perform an analytical study of the propagation of the plane monochromatic waves, by expanding the electromagnetic field into a set of spatially harmonic modes, in a similar way as e.g. described in [11].

$$A(x,z) = \sum_{i,j} A_{i,j} e^{ik_{\perp,i}x + ik_{II,j}z} . \quad (2)$$

$\mathbf{k}_{i,j} = (k_{\perp,i}, k_{II,j}) = (k_\perp + iq_\perp, k_{II} + jq_{II})$, $i,j = ...,-1,0,1,...$.

The expansion results in a coupled system for the amplitudes of harmonics:

$$\left(-2k_0 k_{II,j} - k_{\perp,i}^2\right)A_{i,j} + 2mk_0^2 \sum_{k=i\pm 1, l=j\pm 1} A_{k,l} = 0 \quad (3)$$

Solvability of (3) results in transverse dispersion relation (the dependence of the longitudinal component $k_{II}$ on the transverse component $k_\perp$ of the wave-vector of the Bloch mode), which, as calculated numerically, is given in Fig.1. In the absence of the refractive index modulation $m=0$ the formal solution of (3) consist of a set of parabolas (dashed curves in Fig.1) shifted one with respect to another by the reciprocal vectors of the PC lattice $\mathbf{q}_{1,2}$. They represent the transverse dispersion curves for uncoupled harmonic components of the expansion (2). In nonparaxial description of light propagation these parabolas are to be substituted by circles. The modulation of the refractive index $m \neq 0$ lifts the degeneracy at the crossing points and gives rise to the band-gaps in spatial wave-number domain (Fig.1). For particular relation among the amplitude of modulation given by $m$, for particular geometry of PC given by $\mathbf{q}_{1,2}$, and for a particular frequency given by $k_0 = \omega_0/c$, the plateaus on the transverse dispersion curves appear, indicating the vanishing of diffraction. The inset in Fig.1 indicates the vicinity of the zero diffraction point curve at $k_\perp = 0$.

We next perform an asymptotic analytical description of the diffraction curves at the zero diffraction point by considering only three most relevant expansion modes in (2), those with $(i,j) = (0,0), (-1,-1), (1,1)$. We introduce a set of adimensional parameters by the following normalization for the wavenumbers of light: $K_\perp = k_\perp/q_\perp$, $K_{II} = 2k_{II}k_0/q_\perp^2$. The space coordinates are then rescaled as $Z = zq_\perp^2/2k_0$ and $X = xq_\perp$. Two significant parameters remain after the normalization: $f = 2mk_0^2/q_\perp^2$ which represents the modulation depth of the Bloch mode in the PC, and $Q_{II} = 2q_{II}k_0/q_\perp^2$ which is proportional to the angle between the crystallographic axes of the PC (geometry parameter). The asymptotic solution of (3) in terms of normalized parameters is:

$$K_{II} = K_\perp^2 \left(\frac{8f^2}{(1-Q_{II})^3} - 1\right) + \frac{2f^2}{(1-Q_{II})} \quad (4)$$

The following smallness conditions are to be fulfilled for parameters: $1-Q_{II} = O(\varepsilon)$ and $f = O(\varepsilon^{3/2})$ for the validity of (4). The smallness condition for $K_\perp$ does not need to be fixed at this stage.

The expression for the zero diffraction curve follows directly from (4) eliminating the dependence of $K_{II}$ on transverse wave-vector $K_\perp$: $8f_0^2 = (1-Q_{II,0})^3$. In terms of initial variables it reads:

$$8(2mk_0^2/q_\perp^2)^2 = (1 - 2q_{II}k_0/q_\perp^2)^3 \quad (5)$$

i.e. relates the parameters of the PC $(m, q_{II}, q_\perp)$, and the frequency of monochromatic wave.

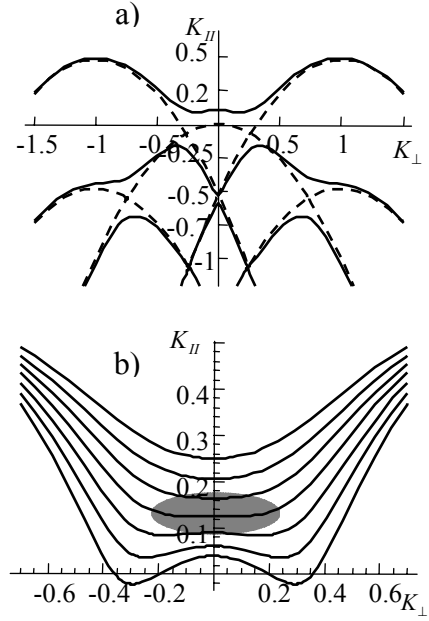

**FIG.1.** The transverse dispersion curve as obtained by numerical solution of (3) using the expansion in 5 harmonics. a) all five dispersion curves (corresponding to five families of eigenmodes), for $f = 0$ (dashed) and $f = 0.15$ (solid). b) upper dispersion curve for different values of the frequency $\delta\omega = -0.2, -0.1, 0, 0.1, 0.2$ from the bottom to upper curve for $f_0 = 0.2$. PC geometry parameter: $Q_{II} = 0.479$

Next we consider a small variation of the carrier frequency $\omega = \omega_0(1+\delta\omega)$, resulting in the variation of the normalized variables in (4): $Q_{II} = Q_{II,0}(1+\delta\omega)$, and $f \to f_0(1+\delta\omega)^2$. Retaining smallness conditions for $1-Q_{II}$ and $f$ as in (4), imposing the following smallness conditions for space and time frequencies: $K_\perp = O(\varepsilon^{3/2})$, and $\delta\omega = O(\varepsilon^2)$, and collecting the terms in (4) up to order $O(\varepsilon^4)$, we obtain:

$$K_{II} = K_{II,0} + \frac{\delta\omega}{V_0} + \frac{\delta\omega^2}{4} + \alpha \cdot \delta\omega K_\perp^2 \quad (6)$$

here $K_{II,0} = (1-Q_{II,0})^2/4$, $1/V_0 = (1-Q_{II,0})\cdot(3-Q_{II,0})/2$, and $\alpha = 3/(1-Q_{II,0})$. The different r.h.s. terms mimic the different ingredients appearing in the PC: a constant shift of the longitudinal wavenumber (term 1) of order of $O(\varepsilon^2)$, a change (decrease) of the group velocity (term 2) of order of $O(\varepsilon^3)$, a group velocity dispersion of the "normal" sign (term 3), and the diffraction dependence on the frequency in the vicinity of the zero diffraction point (term 4). The last two terms appear at the order of $O(\varepsilon^4)$, and is an exceptional property of materials with vanishing diffraction (PCs in this case) close to the inflection point.

We note that different scalings for $K_\perp$ and $\delta\omega$, corresponding to different relations between duration of pulses and width of beam generate (not essentially) different dispersion relation (6). We chose the particular scalings in order to derive the equation (6) with the terms following from normal form analysis at the vicinity of inflection point [12].

Finally, rewriting (6) in terms of time and space variables ($\partial/\partial Z \leftrightarrow iK_{\parallel}$, $\partial/\partial X \leftrightarrow iK_{\perp}$, $\partial/\partial T \leftrightarrow -i\delta\omega$, where the adimensional space variables are as normalized above and the adimensional time is $T = \omega_0 t$) one obtains:

$$\left( \frac{\partial}{\partial Z} - iK_{\parallel,0} + \frac{1}{V_0}\frac{\partial}{\partial T} + \frac{i}{4}\frac{\partial^2}{\partial T^2} - \alpha \frac{\partial}{\partial T}\frac{\partial^2}{\partial X^2} \right) A = 0 \qquad (7)$$

Under approximations used to derive (6), (7) the $A(X,Z,T)$ can be considered as the slow (in space and time) envelope of the Bloch function corresponding to the branch of zero diffraction at zero diffraction point [12].

Next we analyzed the propagation of the Gaussian pulses solving numerically (7). In fact due to the linear character of (7) we calculated the shape of the pulse in spatio-temporal Fourier domain, solved analytically (6) and recovered numerically the shape in space-time domain of propagated pulse. Series of typical results are given in Fig.2:

In general we obtain an expected result that the Gaussian beams disshape during the propagation, due to the dependence of diffraction coefficient on the frequency (term 4 in he r.h.s. (7)).

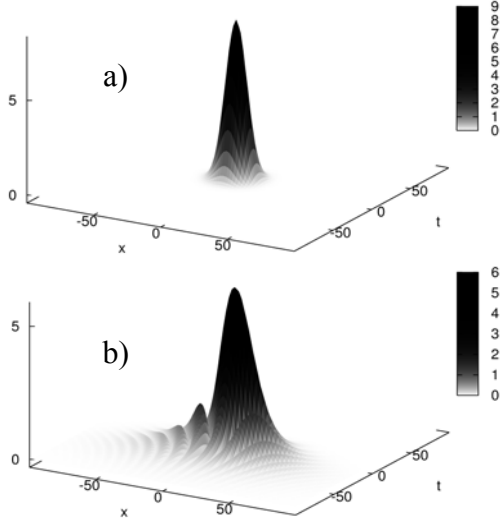

**FIG.2.** Propagation of the Gaussian pulse centered around the zero-diffraction point, as obtained by numerical integration of (7) with $Q_{\parallel,0} = 0.9$: initial Gaussian shape (a), and the pulse propagated over $Z = 15$ (b). The width of the initial Gaussian pulse in Fourier domain is $\Delta K_{\perp} = \Delta \delta\omega = 0.2$.

The overall spatial broadening of the pulsed beam is evaluated by analyzing the propagation of independent frequency components of the radiation, and by averaging over frequency. The result is that the broadening is governed by the efficient diffraction coefficient $d_{eff} \approx d \cdot \Delta\delta\omega$, with the proportionality coefficient close to one, and depending on the location on the zero diffraction curve (to be analyzed in detail elsewhere).

We searched for particular spatio-temporal shapes of the pulses which are invariant under the propagation described by (7). The pulse propagating invariantly with the group velocity $V_1$ is obtained directly from (6):

$$K_{\parallel,1} + \frac{\delta\omega}{V_1} = K_{\parallel,0} + \frac{\delta\omega}{V_0} + \frac{\delta\omega^2}{4} + \alpha\delta\omega K_{\perp}^2 \qquad (8)$$

Here $K_{\parallel,1}$ is a free parameter (having sense of longitudinal wave-number) spanning the one-parameter family of the pulses propagating with a given group velocity $V_1$. Fig.3. shows the families of the isolines of $K_{\parallel,1}$ in the plane of ($\delta\omega, K_{\perp}$) obtained for two different group velocities. The radiation modes within a given isoline propagate with the same wave-vector $K_{\parallel,1}$, therefore the field formations belonging to each separate isoline propagate without dispersive/diffractive broadening.

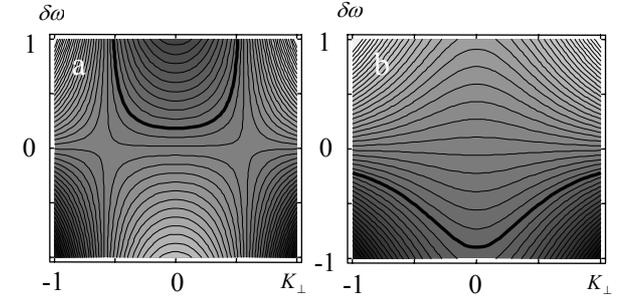

**FIG.3.** Isolines $K_{\parallel} = const$ as calculated from (8) for given values of the velocity: $\Delta V = (1/V_0 - 1/V_1) = -10$ (a), and $\Delta V = +10$, (b) for $Q_{\parallel,0} = 0.9$. The bold isolines are used for preparation of propagation invariant pulses in subsequent Figs.

We note that so called X-pulses are propagation invariant structures residing on hyperbolas in the space-time Fourier domain [9]. Here differently from [9] we obtain a big variety of invariant structures that are mostly of bell shape in space-time Fourier domain (Fig.3). We generate several typical invariant structures. For this purpose we fix the desired frequency in (7), select a particular isoline $K_{\parallel} = const$ (i.e. one isoline from Fig.3.) and generate a spatio-temporal spectrum of the pulse around a selected isoline. The selection of the infinitely narrow spectra $\delta K_{\parallel} \to 0$ results in tails of the pulse expanding to infinity. Therefore we generated the spatio-temporal spectrum of the finite $\delta K_{\parallel}$ centered around a particular isoline. We perform the inverse Fourier transformation and obtain the spatio-temporal shapes of such propagation invariant pulses as shown in Fig.4.a,b. The finite spectrum on one hand results on finite in space and time pulses, on the other hand results in weak spreading of the pulses during the propagation The distance at which the pulses spread sensibly (an analogue of the Reyleigh length) can be evaluated by assuming the spectral components dephase during that propagation: $\delta K_{\parallel} Z \approx 2\pi$, and for pulses shown in Fig.4 is $Z_{Rel} \approx 60$.

Finally we demonstrate the relatively invariant propagation of the pulses. We prepare the pulses as described above, corresponding to those in Fig.4.a,b, and superimpose the both linearly (Fig.4.c). We integrate the evolution equation (7) numerically with a purpose to check simultaneously two phenomenal: 1) the separation of the pulses due to the different velocities fixed in preparation the constituent pulses; 2) the weak spreading of the individual pulses due to the finite width of their spectra. The result of this check is summarized in Fig.4.c where a clear separation of the central peaks of the pulses is visible in propagation, while the shapes of the individual pulses distort negligibly.

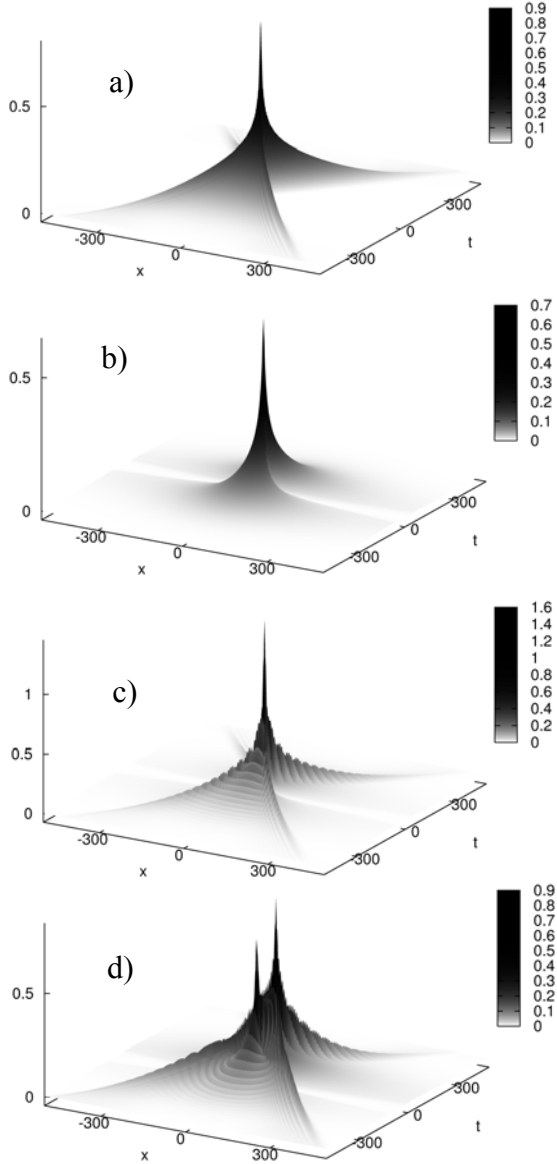

**FIG.4.** Two invariant pulses as constructed by selecting radiation around the bold isolines from Fig.3: a) $V = -10$, $K_{II} = 8.01$, $\delta\omega = 0.2$, $K_\perp = 0.25$; b) $V = 10$, $K_{II} = 5.81$, $\delta\omega = -0.7$, $K_\perp = 0.25$; c) Superposition of pulses a) and b); d) the superposition pattern propagated over the distance $Z = 6$, by solving numerically (7). We note that the fringes appear both in the initial superposition pattern as well as in propagated pattern, which is due to nonuniform phases of the constituent pulses. The "thickness" of the isoline is $\delta K_{II} = 0.1$;

Finally we evaluate the real-world parameters corresponding to the nonspreading pulses. One set of the parameters from a parameter family resulting to a particular set of normalized parameters as used in Fig.4 is the following: $\lambda = 1\,\mu m$, $q_\perp = 0.5 \cdot k_0$, $q_{II} = 0.11 \cdot k_0$ ($\lambda_\perp = 2\,\mu m$ and $\lambda_{II} = 9\,\mu m$ respectively), $m = 0.00125$. Then the evaluated values of the pulses propagating without spreading through PC are: the width of the beam $x_0 \approx 10\,\mu m$, and the duration of the pulse $\tau_0 \approx 17\,fs$.

The work was financially supported by project FIS2004-02587 of the Spanish Ministry of Science and Technology.